# RF Cavity Design

*E. Jensen*
CERN, Geneva, Switzerland

**Abstract**
After a short overview of a general approach to cavity design, we sketch the derivation of waveguide modes from plane waves and of cavity fields from waveguide modes. The characteristic parameters describing cavities and their performance are defined and explained. An equivalent circuit is introduced and extended to explain beam loading and higher order modes. Finally travelling- and standing-wave multi-gap cavities are introduced using the Brillouin diagram.

## 1  Introduction—the outline of a general cavity design procedure

The design of an accelerating cavity or—more generally—an accelerating structure is a complex task, and a one-fits-all recipe cannot be given. Normally the design starts from the *specification* stating the desired performance of the cavity. At the same time, there are some *constraints* which limit the possibilities of implementation. The next issue to consider is the *side effects* the cavity will have, e.g., due to the impedance it presents to the beam at the spurious higher order mode frequencies. These possible undesired effects of the cavity must be taken into account at an early stage of the design. Other parameters, which may be neither explicitly specified nor limiting, should be chosen such as to *optimize* the overall design.

The distinction between 'specifications', 'constraints', 'side effects', and 'optimization parameters' is somewhat arbitrary. Depending on the purpose of the cavity, constraints may become a major part of the specifications, and side effects may be dominating and eventually limiting. Over the last decades, cavity designers have often invented ways to surmount seemingly difficult limits. In this sense, a lecture on 'cavity design procedures' is not meant to narrow down approaches, but rather to enlarge the view.

### 1.1  Purpose of the cavity

Of major importance is the knowledge of what purpose the cavity will serve. This could be acceleration, but also bunching, debunching, or more general 'RF gymnastics' in a synchrotron. The cavity could serve as a 'harmonic cavity' for bunch shaping or to introduce a synchrotron frequency spread to make internal oscillations incoherent, thus increasing stability. In a linac, its purpose could be to get the beam to large energies in the shortest possible distance (to create a larger accelerating gradient) or to give maximum power to the beam. It could, in the case of an RF quadrupole, intentionally cause transverse focussing—other special cavities can serve as transverse kickers.

Other important questions are (1) whether operation is planned for single bunches or long trains of bunches, (2) whether the RF system and the cavity are to operate in continuous wave (CW) or in pulsed mode, (3) how large is the beam current and what is its spectrum—this will determine the requirements on the required power and the dimensioning of the power coupler and the higher order mode couplers and dampers. Once these questions are answered, it is certainly a good idea to find out whether a cavity for a similar range of parameters has already been studied and operated—at least as a starting point from which to begin optimization.

## 1.2 Operation frequency

Normally the frequency is specified; it ranges typically from a few MHz to some GHz. The corresponding wavelengths, from roughly tens of metres down to the centimetre range, indicate already that entirely different physical sizes will result, at least if for illustration one looks at vacuum cavities, which have typical diameters of the order of the wavelength $\lambda$ (the inner diameter of a pillbox cavity, e.g., is $\approx 0.765\,\lambda$, see Eq. (18)). Very large cavities of course become bulky, heavy, and more difficult to keep mechanically stable—for very small cavities tolerances may become unacceptably demanding, and also the size of the beam tube and the power couplers may become impractically small.

For low energy synchrotrons, the frequency is required to sweep over a wide range to compensate for the variation of the particle velocity—this sweep range sometimes spans two octaves. This can be implemented with ferrites inside the cavity volume, tuning to the required frequency by biasing its magnetization with a d.c. magnetic field to vary $\mu$. Standard ferrites typically work only up to 10 MHz; above this frequency so-called microwave ferrites can be used. Another possibility is the use of a wideband cavity based on magnetic alloys such as Finemet®, which can cover such a frequency range instantaneously, i.e., without tuning.

## 1.3 Bandwidth

For special applications, e.g. non-adiabatic RF gymnastics in circular machines, a fast (compared to the synchrotron frequency) variation of the accelerating field is required; there are examples where the RF has to be 'switched' on or off (or by 180° in phase) in less than one RF period—this corresponds to an extremely large bandwidth. A large bandwidth of the system is also needed if fast RF feedback is to be used. This normally concerns the coupling to the amplifier more than it concerns the cavity as such; in particular, for superconducting cavities with a typical $Q_0$ of the order of some $10^{10}$, corresponding to a natural bandwidth of Hz or below, the coupling reduces the loaded $Q$ by many orders of magnitude, thus giving the cavity a manageable bandwidth. Given a maximum available power from the amplifier chain, the bandwidth is limited by the time it takes to store the electromagnetic energy in the cavity—e.g., if 100 J are stored in the cavity at nominal voltage, it takes 100 kW to switch the cavity on in 1 ms.

## 1.4 Tuning

For a narrowband cavity, it must be possible to tune the cavity resonance frequency to the operation frequency: sensitivity of the resonance frequency to unavoidable construction tolerances, or to temperature or pressure variations, may require a tuner to compensate for these effects. For superconducting cavities, the *Lorentz* force detuning, i.e., the detuning due to the electromagnetic field radiation pressure, must not be ignored. For a bunching cavity, where the beam current is in quadrature to the gap voltage, varying the beam current components (see Section 3.10.2) at the RF is equivalent to a detuning, and a fast tuner may have to be envisaged to compensate for them.

## 1.5 Accelerating voltage

The accelerating voltage of a cavity is normally requested to be as high as possible. Bunching and debunching however often also require a stable operation at low voltages, and in particular smooth control over a wide dynamic range from very low to very high voltages. This may be difficult to achieve, in particular if multipacting resonances occur in the operating range.

## 1.6 Beam loading

With increasing beam current, a beam induced voltage will increasingly affect the accelerating voltage. Whilst reactive beam loading leads to detuning, the in-phase component of the beam current will result in resistive beam loading, i.e., it will reduce the generator-excited accelerating voltage. Due to this reduction, the term has a negative connotation, but beam loading is essential for a reasonable power transfer from the RF to the beam; the power transferred to the beam is voltage × beam current, so no power can be extracted from the RF to the beam with zero beam loading. To optimize power transfer efficiency, full account must be taken of beam loading.

The beam thus 'loads' the cavity, and, as explained, it does so at the design frequency. At the same time, however, the beam current also excites higher order modes (HOMs) in the cavity, both longitudinal and deflecting modes, which in return exert forces back on the beam. This may lead to beam instabilities and eventually to beam loss. The cavity also extracts power from the beam, and this power will be lost in higher order mode dampers, which have to be designed to cope with this power.

In the frequency domain, this mechanism is characterized by the beam impedance; in the time domain, it is characterized by the wake functions. This may be considered an unwanted side effect of the cavity, but in fact it is often limiting and may thus even become a main design criterion.

## 1.7 Limitations and constraints

Next we consider the limitations and constraints. A vacuum cavity at low frequencies may, for example, become too large to fit in the beam tunnel; also, the overall length of the cavity is usually limited to the size of an available straight section in the circular machine. The mechanical implementation must be considered early in the design phase. At the high end of the frequency scale, in the mm-wave range, where demanded tolerances are in the μm range or even lower, technology may become a limiting factor. The maximum attainable gap voltage is often not limited by the available power, but rather by field emission or even high voltage breakdown. For superconducting cavities, typically the tangential magnetic field at the surface (equal to the surface current density) is limited; equally, the overall dynamic heat load at cryogenic temperature is often constraining.

## 1.8 Optimization

Factors not included thus far should be used for optimization. The weighting function will depend on the actual case. But we can already say that these factors may include cost (both construction and exploitation), reliability, RF to beam efficiency, operability, maintainability, flexibility, sensitivity to tolerances, and/or others.

The beam–cavity interaction is generally described by a few figures of merit, which will be treated in more detail in the following. Besides the operating frequency, these are the shunt impedance and the $R/Q$, both for the fundamental mode and the higher order modes.

## 2 Field–particle interaction

In classical electrodynamics, the interaction between electromagnetic fields and moving charged particles is described by two sets of equations: Maxwell's equations describing the fields and the equations of motion describing the particle dynamics. The moving charged particles will appear as sources $\vec{J}$ and $\rho$ in Maxwell's equations, and the fields will appear as the *Lorentz* force in the equations of motion.

Maxwell's equations in vacuum read as

$$
\begin{aligned}
\nabla \times \vec{B} - \frac{1}{c^2}\frac{\partial}{\partial t}\vec{E} = \mu_0 \vec{J}, &\qquad \nabla \cdot \vec{B} = 0, \\
\nabla \times \vec{E} + \frac{\partial}{\partial t}\vec{B} = 0, &\qquad \nabla \cdot \vec{E} = \mu_0 c^2 \rho,
\end{aligned}
\tag{1}
$$

and the equations of motion of a charged particle of mass $m$ and charge $q$ read

$$
\begin{aligned}
\frac{d\vec{r}}{dt} &= \frac{\vec{p}}{\gamma m}, \\
\frac{d\vec{p}}{dt} &= q\left(\vec{E} + \frac{\vec{p}}{\gamma m} \times \vec{B}\right),
\end{aligned}
\tag{2}
$$

where the relativistic factor $\gamma$ is given by

$$\gamma = \sqrt{1 + \left(\frac{p}{mc}\right)^2} \tag{3}$$

and the total energy of the particle is

$$W = \gamma mc^2 = \sqrt{(mc^2)^2 + (pc)^2} = mc^2 + W_{\text{kin}}. \tag{4}$$

The change of energy is derived from

$$W\,dW = c^2 \vec{p} \cdot d\vec{p} = qc^2 \vec{E}\,dt. \tag{5}$$

Note that only the electric field performs work on the particle (changes its energy); since the magnetic force is rectangular to the particle trajectory, the term vanishes due to the scalar product.

## 3 From waveguide to cavity

Maxwell's equations, along with the boundary conditions of an empty cavity, have solutions with non-vanishing fields even if no sources are present; these solutions are denoted 'eigensolutions', since they correspond to eigenvectors of a generalized eigenvalue problem. These special solutions have harmonic time dependence—the characteristic frequencies (the eigenfrequencies) correspond to the eigenvalues. Without the rigour of a full algebraic derivation, we will sketch here some elements required in this transformation of Maxwell's equations that illustrate this line of thought.

With vanishing currents and charges, Maxwell's equations become

$$\begin{aligned}\nabla \times \vec{B} - \frac{1}{c^2}\frac{\partial}{\partial t}\vec{E} &= 0, & \nabla \cdot \vec{B} &= 0, \\ \nabla \times \vec{E} + \frac{\partial}{\partial t}\vec{B} &= 0, & \nabla \cdot \vec{E} &= 0.\end{aligned} \tag{6}$$

Applying the curl operator ($\nabla \times$) to the third equation results in $\nabla \times \nabla \times \vec{E} + \nabla \times \frac{\partial}{\partial t}\vec{B} = 0$; the time derivative of the first equation yields $\nabla \times \frac{\partial}{\partial t}\vec{B} = \frac{1}{c^2}\frac{\partial^2}{\partial t^2}\vec{E}$. Combining the resulting equations and using the vector identity $\nabla \times \nabla \times \vec{x} \equiv \nabla\nabla \cdot \vec{x} - \Delta\vec{x}$, the *Laplace* equation in four dimensions results:

$$\Delta\vec{E} - \frac{1}{c^2}\frac{\partial^2}{\partial t^2}\vec{E} = 0. \tag{7}$$

### 3.1 Rectangular waveguide

One possible solution of Eq. (7) is a homogeneous plane wave at a fixed frequency, fully described by the solution

$$\vec{E} \propto \vec{u}_y \cos(\omega t - \vec{k} \cdot \vec{r}), \tag{8}$$

where the vector $\vec{k}$ points in the direction of the propagation of the wave and the length of $\vec{k}$ is $k = \omega/c$ in free space. We have chosen here a linear polarization of the electric field in the $y$-direction and a propagation in the $xz$-plane in a direction $\alpha$ with $\vec{k} \cdot \vec{r} = \frac{\omega}{c}(z \cos\alpha + x \sin\alpha)$. Also we have chosen a time dependence of $\cos(\omega t)$, but it should be clear that by rotation of the coordinate system and a shift of the origin this solution is quite general. The illustration on the left of Fig. 1 shows the amplitude of the electric field and should be imagined moving at the speed of light to the right (in the direction of vector $\vec{k}$), orthogonal to the phase fronts (black lines). In the diagram on the right of Fig. 1 we illustrate the components of the vector $\vec{k}$ along the $z$-axis ($k_z$) and the $x$-axis ($k_\perp$)—note that the propagation along the $z$-axis is described by the $z$-component of $\vec{k}$.

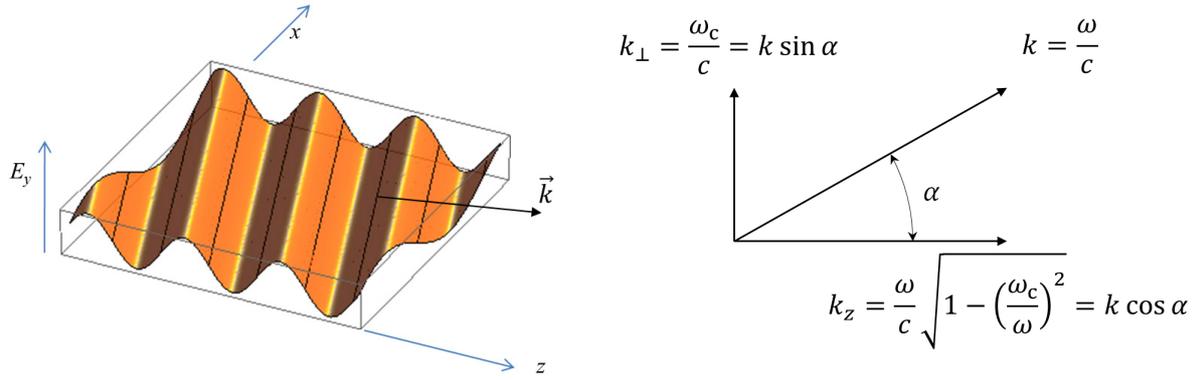

**Fig. 1:** Illustration of a homogeneous plane wave propagating at an angle $\alpha$ w.r.t. the horizontal $z$-axis

Also note that the wavelength along the $z$-axis is $\lambda/\cos\alpha$ and thus larger than the free space wavelength $\lambda$. The velocity of the phase front propagation in a given direction is referred to as the phase velocity, and, since the wave is propagating at speed $c$ in direction $\vec{k}$, the phase velocity in the $z$-direction from the above example is given by

$$v_{\varphi,z} = \frac{\omega}{k_z} = \frac{c}{\cos\alpha} = \frac{c}{\sqrt{1-\left(\frac{\omega_c}{\omega}\right)^2}} > c. \tag{9}$$

The constant $\omega_c = \omega \sin\alpha$ introduced here will be explained below. Since the electric field in the considered example is polarized in the $y$-direction, perfectly conducting planes at any position $y = $ const., i.e., orthogonal to this polarization, can be inserted and will not affect the solution given.

Next consider the superposition of two homogeneous plane waves in the same configuration as above, with equal amplitudes and propagating at angles $\alpha$ and $-\alpha$ w.r.t the $z$-direction. The total field now becomes

$$E_y \propto \left(\cos\left(\omega t - \frac{\omega}{c}(z\cos\alpha + x\sin\alpha)\right) + \cos\left(\omega t - \frac{\omega}{c}(z\cos\alpha - x\sin\alpha)\right)\right)$$
$$= 2\cos\left(\frac{\omega}{c}\sin\alpha\ x\right)\cos\left(\omega t - \frac{\omega}{c}\cos\alpha\ z\right) = 2\cos(k_\perp x)\cos(\omega t - k_z z), \tag{10}$$

from which one can see that in some planes $x = $ const.; e.g., at $\frac{\omega}{c}\sin\alpha\ x = k_\perp x = \pm\frac{\pi}{2}$, the electric field is vanishing at all times. Figure 2 illustrates this superposition. We can now also add perfectly conducting walls at these distinct positions without perturbing the field distribution between them, thus creating a rectangular waveguide of width $a = \frac{\pi c}{\omega_c} = \frac{\pi}{k_\perp}$ and arbitrary height.

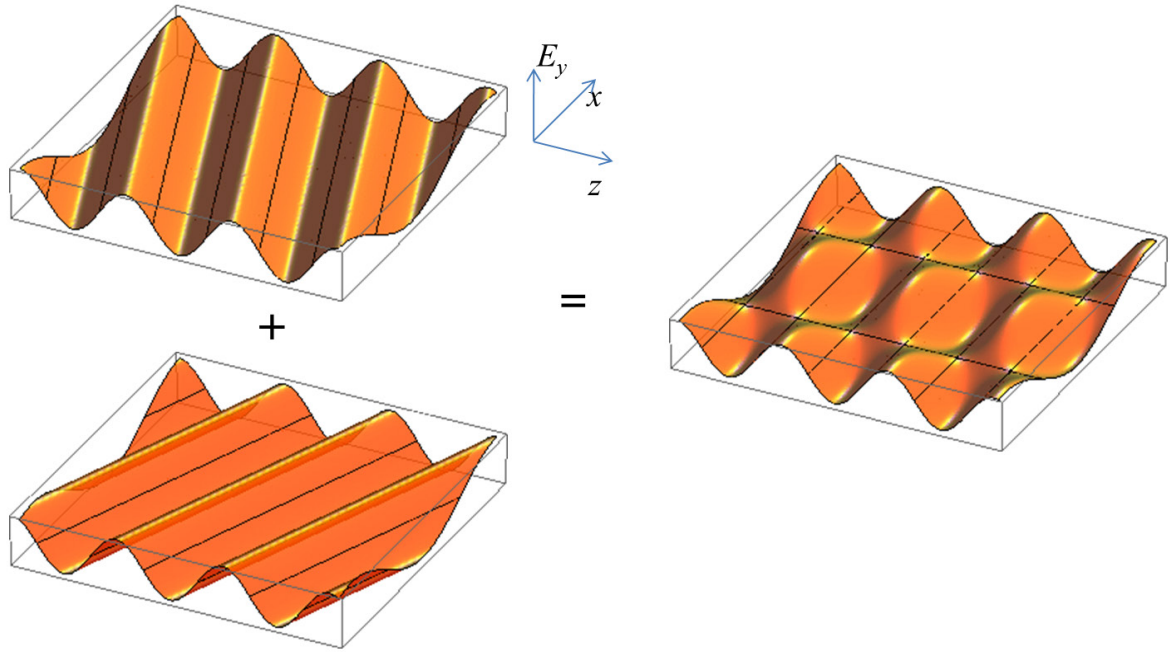

**Fig. 2:** The superposition of two homogeneous plane waves forms a waveguide mode

There exist other locations where perfectly conducting walls can be placed, and other combinations of homogeneous plane waves can be thought of as 'constructing' waveguide modes that satisfy the boundary conditions of the rectangular hollow waveguide; the presented mode is a special case selected for illustration referred to as the TE$_{10}$ mode, where TE refers to 'transverse electric' (no $z$-component of the electric field) and the index '10' (read 'one zero') refers to the orders of the mode in the $x$- and $y$-directions (one half-wave in the $x$-direction, no field dependence on $y$).

### 3.2 Waveguide dispersion–phase velocity

We start with such a waveguide of width $a$; let us vary the frequency $\omega$ and look at possible field distributions inside the waveguide. The given width $a$ now freezes the $x$-component of the vector $\vec{k}$, i.e., $k_\perp$ must now satisfy the condition $k_\perp a = m\pi$ for a TE$_{m0}$ mode; this results in a propagation constant in the $z$-direction of

$$k_z = \sqrt{k^2 - k_\perp^2} = \frac{\omega}{c}\sqrt{1 - \left(\frac{\omega_c}{\omega}\right)^2}, \tag{11}$$

which is the characteristic waveguide dispersion plotted in Fig. 3. Note that the phase velocity along the $z$-axis is given by Eq. (9). At the frequency $\omega_c$, the so-called cutoff frequency, $k_z = 0$ and $k_\perp = k$, which means that the electromagnetic wave bounces off the side walls of the waveguide in a purely transversal direction, at all locations $z$ in phase, i.e., the phase velocity $v_{\varphi,z}$ is infinite; the group velocity, describing the transport of field energy, is zero at cutoff. For $\omega < \omega_c$, $k_z$ is imaginary—there is no wave propagation, but there is exponential field attenuation in the axial direction.

We have introduced $\omega_c$ already in Fig. 1 and Eq. (9)—now its meaning should become clear. The more general expression for $\omega_c$ for a mode of order $m, n$ (TE or TM) is

$$\frac{\omega_c}{c} = \sqrt{\left(\frac{m\pi}{a}\right)^2 + \left(\frac{n\pi}{b}\right)^2}. \tag{12}$$

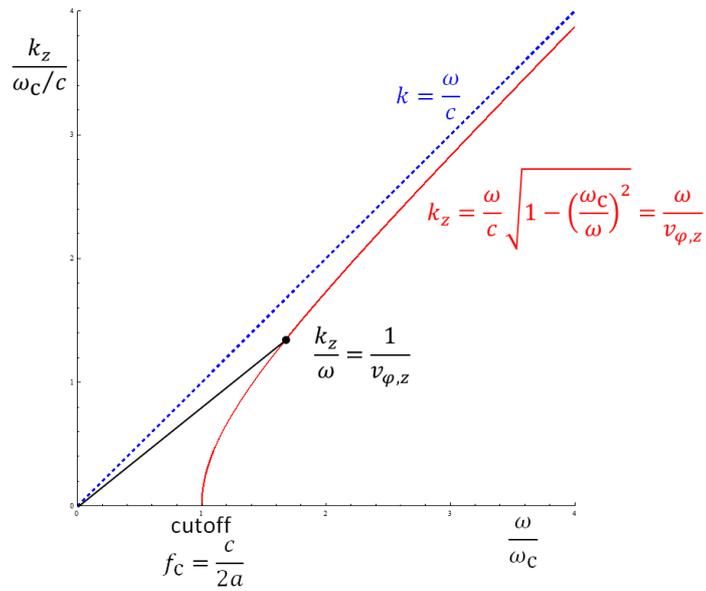

**Fig. 3:** Waveguide dispersion (red solid curve) compared to the free space dispersion (blue dashed curve). The phase velocity is the inverse of $k_z/\omega$; it is infinite at cutoff and approaches $c$ asymptotically for large $\omega$. The curve is normalized to cutoff and is thus identical for all waveguide modes.

To illustrate the field distributions for modes of arbitrary order in a rectangular waveguide, the transverse field distributions of a number of TE and TM modes is rendered in Fig. 4 using arrows and colour coding.

As can be seen, the indices $m$ and $n$ indicate the number of half-waves that fit transversally in the width $a$ and height $b$ of the waveguide, respectively.

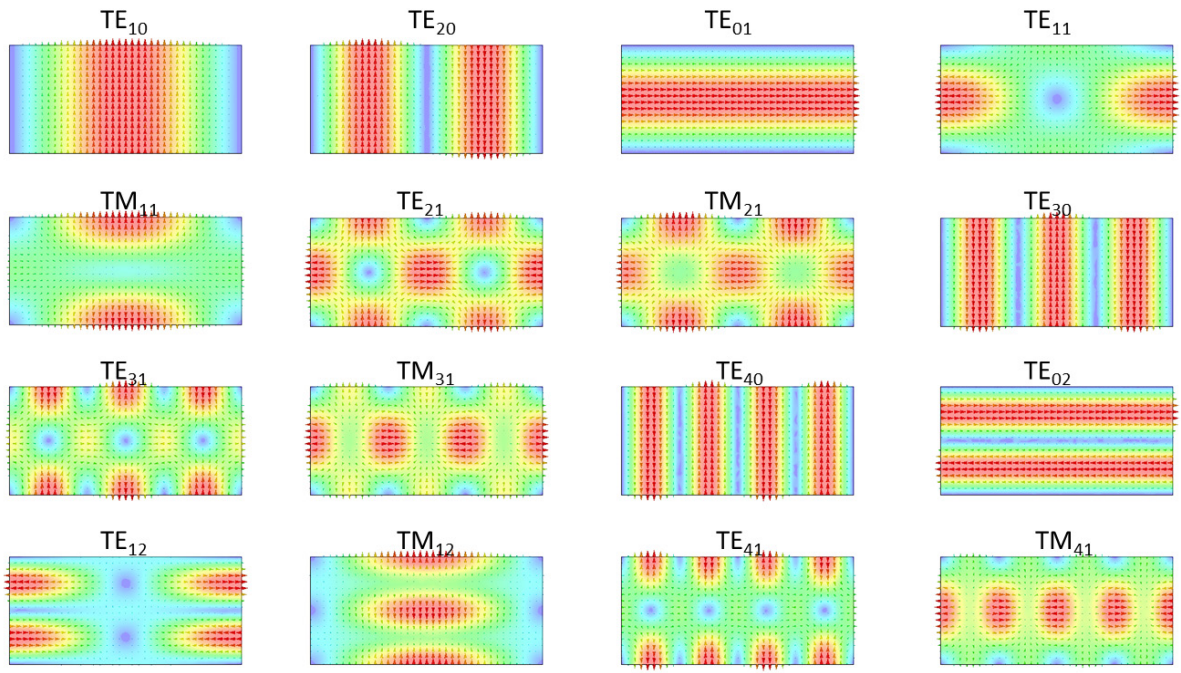

**Fig. 4:** Transverse electric field distributions of different modes in a rectangular waveguide

## 3.3 Round waveguides

The superposition of homogeneous plane waves also allows—at least conceptually—the 'construction' of modes in round waveguides. Generalizing Eq. (10) to a summation over infinitely many directions $\alpha$, i.e., transforming it to an integral $\frac{1}{2\pi}\int d\alpha$, and expressing the transverse position in cylindrical coordinates, $x = \rho \sin\vartheta, z = \rho \cos\vartheta$, we obtain an expression for the resulting field $E_y$ (attention: here we understand $z$ and $x$ as transverse coordinates and $y$ as the axial direction, to derive the field of the circular TM$_{01}$ mode) as

$$\frac{1}{2\pi}\int_0^{2\pi} \cos(k\rho(\cos\alpha\cos\vartheta + \sin\alpha\sin\vartheta))\,d\alpha$$
$$= \frac{1}{2\pi}\int_0^{2\pi} \cos(k\rho(\cos(\alpha-\vartheta)))\,d(\alpha-\vartheta) = J_0(k\rho), \tag{13}$$

which is the integral representation of the Bessel function $J_0$. Similar to Fig. 2, the result of this superposition is illustrated in Fig. 5. We now obtain a radial standing-wave pattern, described by the Bessel function $J_0$. Thanks to the zeros of $J_0$ there exist radial positions where the electric field is always zero. At these radii, perfectly conducting boundary conditions may again be inserted without perturbing this field distribution—we have constructed a circular waveguide!

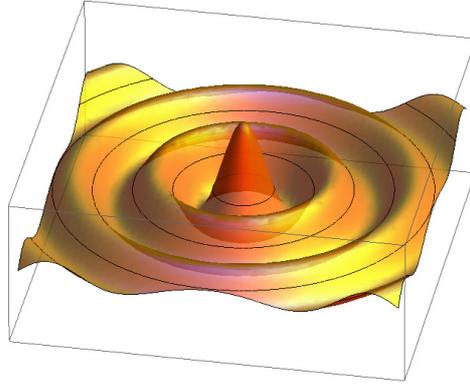

**Fig. 5:** Bessel function $J_0(k\rho)$, which can be obtained by superimposing homogeneous plane waves. Metallic walls may be inserted where $J_0(k\rho) = 0$ (black lines).

The cutoff frequencies of TM modes in a round waveguide of radius $a$ are

$$\frac{\omega_c}{c} = \frac{\chi_{m,n}}{a}, \tag{14}$$

where $\chi_{m,n}$ denotes the $n$th zero of the Bessel function $J_m$, and the cutoff frequencies of the TE modes are

$$\frac{\omega_c}{c} = \frac{\chi'_{m,n}}{a}, \tag{15}$$

with the $n$th zero of the derivative of the Bessel function $J_m$. The lowest order mode in a round waveguide is the TE$_{11}$ mode, for which $\chi'_{11} \approx 1.84118$; the lowest order TM mode is the TM$_{01}$ mode (which we used for the introduction) with $\chi_{01} \approx 2.40483$.

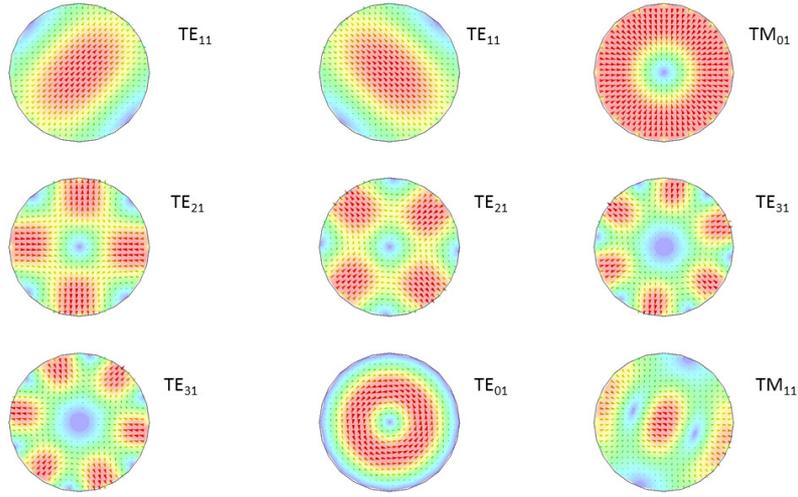

**Fig. 6:** Transverse electric-field distributions of different modes in a round waveguide

**Table 1:** Collection of parameters and formulae describing waveguide modes

| General cylindrical waveguide | TE modes | TM modes |
|---|---|---|
| Boundary conditions | $\vec{n} \cdot \nabla T = 0$ | $T = 0$ |
| Longitudinal wave equations | $\dfrac{dU(z)}{dz} + jk_z Z_0 I(z) = 0, \quad \dfrac{dI(z)}{dz} + \dfrac{jk_z}{Z_0} U(z) = 0$ | |
| Propagation constant | $k_z = k\sqrt{1 - \left(\dfrac{\omega_c}{\omega}\right)^2}$ | |
| Characteristic impedance | $Z_0 = \dfrac{\omega \mu}{k_z}$ | $Z_0 = \dfrac{k_z}{\omega \varepsilon}$ |
| Ortho-normal eigenvectors | $\vec{e} = \vec{u}_z \times \nabla T$ | $\vec{e} = -\nabla T$ |
| Transverse fields | $\vec{E} = U(z)\,\vec{e}, \quad \vec{H} = I(z)\,\vec{u}_z \times \vec{e}$ | |
| Longitudinal fields | $H_z = \left(\dfrac{\omega_c}{\omega}\right)^2 \dfrac{T\,U(z)}{j\omega\mu}$ | $E_z = \left(\dfrac{\omega_c}{\omega}\right)^2 \dfrac{T\,I(z)}{j\omega\varepsilon}$ |
| **Rectangular waveguide** (width $a$, height $b$) (with $\varepsilon_i = \begin{cases}1 & \text{if } i=0 \\ 2 & \text{if } i \neq 0\end{cases}$) | | |
| Cutoff | $\dfrac{\omega_c}{c} = \sqrt{\left(\dfrac{m\pi}{a}\right)^2 + \left(\dfrac{n\pi}{b}\right)^2}$ | |
| Transverse eigenfunction $T_{mn}$ | $\dfrac{1}{\pi}\sqrt{\dfrac{a\,b\,\varepsilon_m \varepsilon_n}{(mb)^2+(na)^2}} \cos\left(\dfrac{m\pi}{a}x\right)\cos\left(\dfrac{n\pi}{b}y\right)$ | $\dfrac{2}{\pi}\sqrt{\dfrac{a\,b}{(mb)^2+(na)^2}} \sin\left(\dfrac{m\pi}{a}x\right)\sin\left(\dfrac{n\pi}{b}y\right)$ |
| **Round waveguide (radius $a$)** | | |
| Cutoff | $\dfrac{\omega_c}{c} = \dfrac{\chi'_{mn}}{a}$ | $\dfrac{\omega_c}{c} = \dfrac{\chi_{mn}}{a}$ |
| Transverse eigenfunction $T_{mn}$ | $\sqrt{\dfrac{\varepsilon_m}{\pi(\chi'^2_{mn}-m^2)}}\,\dfrac{J_m\left(\chi'_{mn}\tfrac{\rho}{a}\right)}{J_m(\chi'_{mn})} \begin{Bmatrix}\cos(m\varphi)\\ \sin(m\varphi)\end{Bmatrix}$ | $\sqrt{\dfrac{\varepsilon_m}{\pi}}\,\dfrac{J_m\left(\chi_{mn}\tfrac{\rho}{a}\right)}{J_{m-1}(\chi_{mn})} \begin{Bmatrix}\sin(m\varphi)\\ \cos(m\varphi)\end{Bmatrix}$ |

## 3.4 Fields in rectangular and round waveguides

Transverse eigenfunctions $T$ allow us to generalize to waveguides with arbitrary cross-sections; they solve the transverse wave equation (aka the membrane equation): $\Delta T + \left(\frac{\omega_c}{c}\right)^2 T = 0$. In the general case, the longitudinal distributions of the transverse electric and magnetic fields for each mode are described by the functions $U(z)$ and $I(z)$, respectively. $U(z)$ and $I(z)$ are derived from the longitudinal wave equations, which feature the characteristic impedance $Z_0$. Table 1 summarizes the main parameters and formulae to describe modes in cylindrical waveguides.

## 3.5 Standing waves—pillbox cavity

A hollow waveguide—be it round, rectangular, or of any other shape—limits the possible transverse components of the wave vector to some discrete values that 'fit' in the geometry (e.g., $m\pi/a$ and $n\pi/b$ in a rectangular waveguide). This results in a certain axial component of $\vec{k}$, the propagation constant, since the total length of $\vec{k}$ is of course given by the frequency. If a length of waveguide $\ell$ is now also closed by conducting walls at its beginning and its end, forming a closed 'cavity', the axial component also of $\vec{k}$ is confined to discrete values—which will now determine also the length of $\vec{k}$, in other words the frequency. Non-vanishing fields exist in the cavity only for discrete frequencies; they are called eigenfrequencies. For the example of a rectangular waveguide, this results in a *shoebox* shaped cavity with discrete eigenfrequencies $\omega_{m,n,p}$ which are given by

$$\frac{\omega_{m,n,p}}{c} = \sqrt{\left(\frac{m\pi}{a}\right)^2 + \left(\frac{n\pi}{b}\right)^2 + \left(\frac{p\pi}{\ell}\right)^2}. \tag{16}$$

For a TM mode in a circular waveguide, the corresponding equation is

$$\frac{\omega_{m,n,p}}{c} = \sqrt{\left(\frac{\chi_{m,n}}{a}\right)^2 + \left(\frac{p\pi}{\ell}\right)^2}. \tag{17}$$

The name '*pillbox* cavity' is derived from the shape of this cavity for small $\ell$. The fundamental mode of a short ($\ell < 2.03076\, R$) pillbox cavity is the TM$_{010}$ mode, given by $m$:

$$\frac{\omega_{0,\text{pillbox}}}{c} = \frac{\chi_{01}}{a} \approx \frac{2.40483}{a}. \tag{18}$$

The fields of the TM$_{010}$ mode (pillbox) are given by

$$E_z = \frac{1}{j\omega_0 \varepsilon_0} \frac{\chi_{01}}{a} \sqrt{\frac{1}{\pi}} \frac{J_0\left(\chi_{01} \frac{\rho}{a}\right)}{a J_1(\chi_{01})}, \quad B_\varphi = \mu_0 \sqrt{\frac{1}{\pi}} \frac{J_0\left(\chi_{01} \frac{\rho}{a}\right)}{a J_1(\chi_{01})} \tag{19}$$

$m = p = 0$, $n = 1$, where $p = 0$ indicates that there is no axial field dependence—the eigenfrequency is actually the cutoff frequency of the TM$_{01}$ mode of the round waveguide:

Figure 7 shows an idealized pillbox cavity (on the left) along with some more realistic shapes derived from the idealized pillbox.

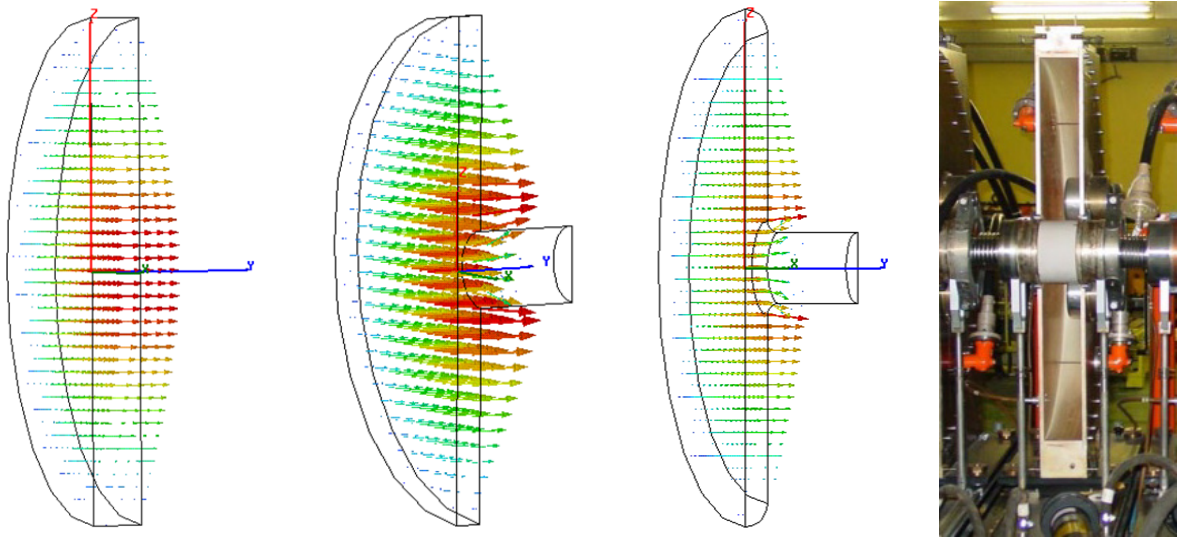

**Fig. 7:** Diagrams from left to right: an idealized pillbox cavity; with added beam pipe opening (quarter shown); with added rounding to reduce field enhancement near sharp edges. These are results of field calculations—the arrows indicate the electric field. Far right: the opened PS 200 MHz cavity with pillbox shape and ceramic gap.

Another way of looking at a length $\ell$ of waveguide short circuited at its two ends is to look at the signal flow chart (Fig. 8): assume a waveguide mode travelling forward, reflected with reflection coefficient −1, travelling backward, and reflected again. If we assume an amplitude $A$ at some location, the resulting equation

$$A = e^{-jk_z 2\ell} A \qquad (20)$$

is an eigenvalue equation for the amplitude $A$. Non-vanishing solutions exist for $e^{-jk_z 2\ell} = 1$, so again for integer multiples $p$ of half-waves that fit the length $\ell$. For each eigenfrequency there is a characteristic field distribution in the cavity; whilst the eigenfrequency corresponds to an eigenvalue, the field distribution corresponds to its eigenvector.

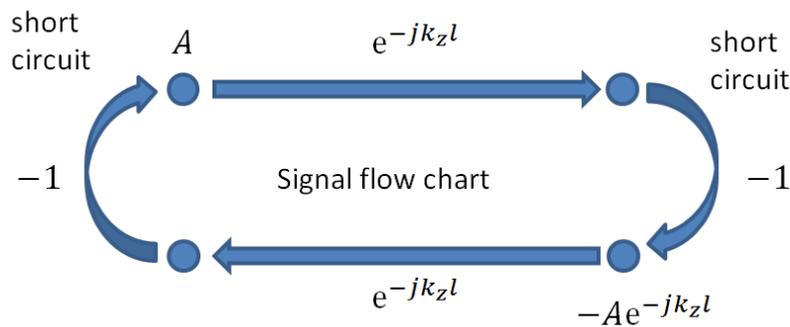

**Fig. 8:** Signal flow chart of a waveguide mode fully reflected at either end of a length of waveguide

### 3.5.1 Re-entrant cavity

Most accelerating cavities of course do not quite look like pillboxes—their shape is typically circular around the beam axis and they feature transitions to the tubes for the incoming and outgoing beams, but otherwise their shapes vary. A sample shape is given in Fig. 9, which indicates the electric-field lines of the fundamental mode; another example is given in Fig. 14. The principle though, that the outer metallic boundary allows only for a discrete, numerable set of eigenfrequencies associated with their characteristic field distributions, remains valid.

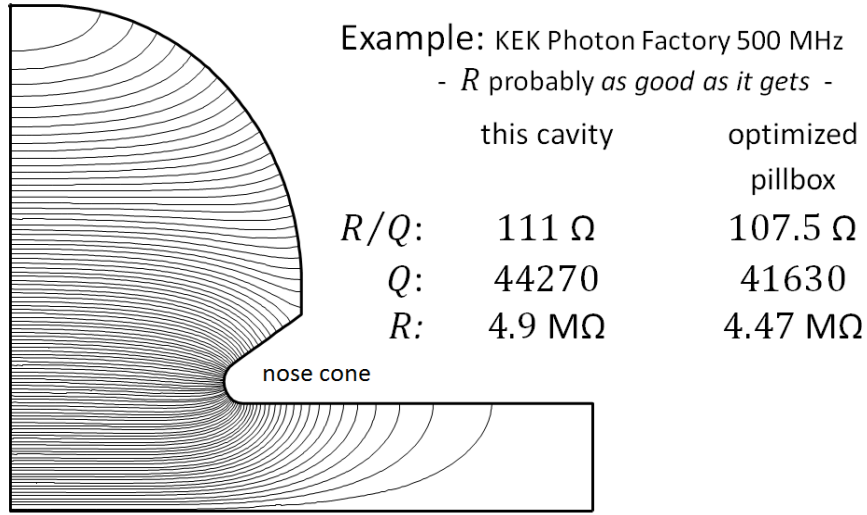

**Fig. 9:** Example of a re-entrant cavity (with nose cones) of optimized shape. Example taken from KEK Photon Factory. Superfish output—only half of the symmetric cavity profile is shown.

The fields in the pillbox cavity can be given in closed form and exact, but in the general case the electromagnetic fields are calculated numerically—a number of specialized simulation programs are commercially available, including, e.g., Superfish [1], HFSS [2], CST Microwave Studio [3], and GdfidL [4].

### 3.6   Accelerating voltage

Once the field distribution (of the considered accelerating mode) is known, the interaction with the charged particle beam can be computed. The accelerating voltage is defined as the integral of the axial electric field along the particle trajectory, taking the finite speed $\beta c$ of the particle into account:

$$V_{\text{acc}} = \int_{-\infty}^{\infty} E_z e^{-j\frac{\omega z}{\beta c}} \, \mathrm{d}z. \tag{21}$$

The exponential term takes care of the fact that the field is varying while the particle is traversing the gap. The integral is extended here from $-\infty$ to $+\infty$ assuming that the beam tube diameter is small enough to form a waveguide well below its lowest cutoff at the resonance frequency $\omega$, such that the fields extending from the cavity into the beam tube are rapidly decaying to zero (as is also clearly visible in Fig. 9). $V_{\text{acc}}$ is generally a complex quantity—this becomes important when many gaps are considered, where the voltages must be added with their correct phases.

#### 3.6.1   Transit time factor

The definition of the accelerating voltage given in the preceding paragraph is not the only possible one—in not every definition is the finite particle speed accounted for. This allows the definition of the so-called transit time factor, defined as

$$TT = \frac{\left|\int_{-\infty}^{\infty} E_z e^{-j\frac{\omega z}{\beta c}} \, \mathrm{d}z\right|}{\int_{-\infty}^{\infty} |E_z| \, \mathrm{d}z}. \tag{22}$$

The transit time factor describes the reduction of the voltage that a real particle experiences relative to a fictitious particle with infinite speed. It is instructive again to look at the simple pillbox cavity with the $z$-independent $E_z$ from Eq. (19), where the expression for $TT$ simply becomes

$$TT_{\text{pillbox}} = \frac{\left|\sin\left(\frac{\chi_{01}\ell}{2a}\right)\right|}{\frac{\chi_{01}\ell}{2a}}, \quad (23)$$

plotted in Fig. 10. Note that $TT = 1$ for very small gaps, and decreases for larger gaps. It reaches zero when the transit time of the charged particle is an entire RF period, i.e., when $\ell = \beta\lambda$.

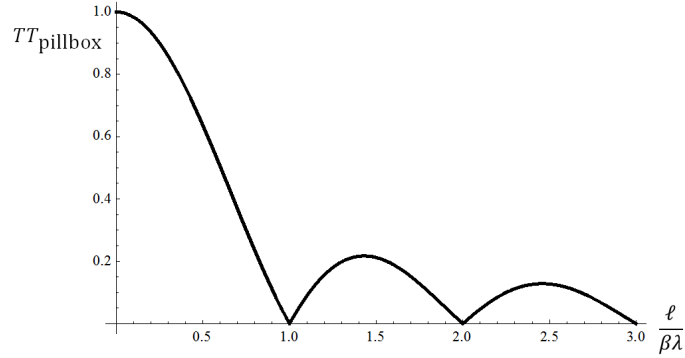

**Fig. 10:** The transit time factor of a pillbox cavity as a function of its length $\ell$

### 3.7 Stored energy

There is energy stored in the electromagnetic field; for a cavity mode, where the electric and magnetic fields are 90° out of phase, the energy is continuously swapping between the magnetic energy and the electric energy such that, on average, the electric and magnetic stored energies are equal. The energy stored in a mode is given by

$$W = W_E + W_M = \iiint \frac{\varepsilon}{2}|E|^2 \, dV + \iiint \frac{\mu}{2}|H|^2 \, dV, \quad (24)$$

where the volume integral is to be extended over the entire cavity volume.

#### 3.7.1 R upon Q

Note that the stored energy is proportional to the square of the accelerating voltage; this allows the definition of the quantity referred to as '$R$ upon $Q$', or $R/Q$, of the considered mode of a cavity as

$$\frac{R}{Q} = \frac{|V_{\text{acc}}|^2}{2\omega_0 W}. \quad (25)$$

$R/Q$ should be considered as a fundamental quantity (even if its notation suggests a derived quantity); it is independent of the cavity losses described in the following and depends only on the cavity geometry.

In the case of the simple, idealized pillbox cavity, the integrals (21) and (24) can be evaluated analytically using the fields from Eq. (19). When evaluating the integral one notes that in fact $W_E = W_M$; the result for $R/Q$ is

$$\frac{R}{Q}\bigg|_{\text{pillbox}} = \frac{4\eta}{\chi_{01}^3 \pi J_1^2(\chi_{01})} \frac{\sin^2\left(\frac{\chi_{01}}{2}\frac{\ell}{a}\right)}{\frac{\ell}{a}}, \quad (26)$$

where $\eta = \sqrt{\frac{\mu_0}{\varepsilon_0}} = 376.73\ \Omega$ is the free space wave impedance.

## 3.8 Wall losses

A number of mechanisms can lead to energy loss in a cavity, but for a vacuum cavity the dominating losses are ohmic losses in the not perfectly conducting walls. Even for superconducting cavities they must not be neglected. But since these losses are normally small by design, a perturbation method (as follows) can be applied to calculate them.

The tangential magnetic field near the metallic surface is equal to a surface current $\vec{J}_A = \vec{n} \times \vec{H}$ (in A/m, describing a current concentrated in the thin 'skin depth'). This surface current sees a surface resistance $R_A = \sqrt{\frac{\omega\mu}{2\sigma}}$ (in Ω), which will cause a small tangential voltage drop and thus resistive losses (an outward normal component of the *Poynting* vector $\vec{S} = \Re\{\vec{E} \times \vec{H}^*\} = \Re\{R_A \vec{H}_t \cdot \vec{H}_t^*\} = \frac{1}{2}R_A|H_t|^2$). To determine the total wall losses, this power density has to be integrated over the entire cavity surface to yield

$$P_{\text{wall}} = \frac{1}{2} \oiint_{\text{wall}} R_A |H_t|^2 \, dA. \tag{27}$$

This is a perturbation method since we are assuming that the effect of the tangential voltage drop can be neglected compared to the dominating normal electric field of the mode.

Note that the surface resistance $R_A$ defined and used above is directly related to the skin depth $\delta$ via the formula

$$\delta \sigma R_A = 1, \tag{28}$$

where $\sigma$ is the conductivity of the surface wall material.

Here we briefly indicate the orders of magnitude of typical conductivities: Cu at room temperature has $\sigma \approx 5.8 \cdot 10^7$ S/m, which leads to $R_A \approx 8$ mΩ at 1 GHz, scaling with $\sqrt{\omega}$. Nb at 2 K has—depending on the technology and frequency—a surface resistance of $R_s \approx 10$ nΩ at 1 GHz, scaling with $\omega^2$. For superconducting cavities, $R_s$ is often used instead of $R_A$.

### 3.8.1 Quality factor

Note that the losses are directly proportional to the stored energy and to the square of the accelerating voltage. This allows the definition of other quantities used to describe cavities (or rather specific modes in cavities)—the first one is relating the power loss to the stored energy and is called $Q$ or the quality factor. The quality factor of a mode relates the stored energy to the energy loss during one RF period:

$$Q = \frac{\omega_0 W}{P}. \tag{29}$$

The larger $Q$ is, the less power will be needed to sustain stored energy inside the cavity.

We have discussed here power loss in the cavity wall, but other loss mechanisms can readily be included in this notation. Losses in dielectric or magnetic materials inside the cavity, losses caused by radiation through openings, or even losses due to discharges inside the cavity may be considered. The power 'lost' through the main power coupler can be included—this latter is called the external $Q$ ($Q_{\text{ext}}$). Since the power lost due to these different loss mechanisms must be added, the total resulting $Q$, also referred to as $Q_L$ (the loaded $Q$), is calculated from

$$\frac{1}{Q_L} = \frac{P_{\text{wall}} + P_{\text{ext}} + \cdots}{\omega_0 W} = \frac{1}{Q_0} + \frac{1}{Q_{\text{ext}}} + \frac{1}{\cdots}, \tag{30}$$

where $Q_0$ denotes the unloaded $Q$, resulting from just the wall losses in our case. Typical $Q_0$ values obtained in normal-conducting vacuum cavities made of Cu are $\mathcal{O}(10^3 \ldots 10^6)$, depending on geometry, size, and frequency; typical values for superconducting cavities are $\mathcal{O}(10^9 \ldots 10^{11})$.

A quantity often used is the coupling factor $\beta$; it is the ratio $P_{\text{ext}}/P_{\text{wall}}$. Consequently the loaded $Q$ of the cavity is reduced from the unloaded $Q$ by a factor $(1 + \beta)^{-1}$:

$$Q_{\text{L}} = \frac{Q_0}{(1 + \beta)}. \tag{31}$$

It will become clear in the following why $\beta$ is called the coupling factor.

### 3.8.2 Shunt impedance

Combining the quantities introduced in Eqs. (25) and (29) results in the relation between the power lost and the accelerating voltage, the so-called shunt impedance:

$$R = \left(\frac{R}{Q}\right) Q = \frac{|V_{\text{acc}}|^2}{2P}. \tag{32}$$

Maximizing the shunt impedance $R$ allows optimization of the accelerating *voltage* that can be obtained for a given available power. $R$ is classically the quantity that is to be optimized; it should be noted however that it may not be the quantity to optimize if the beam current is substantial and will itself excite a voltage in this same impedance (see Section 3.10.2).

### 3.8.3 Geometric factor

Another quantity is used practically only for superconducting cavities—it relates $Q_0$ and the surface resistance $R_A$ or $R_s$ used in Eq. (27) and defines the so-called 'geometric factor' $G$:

$$G = Q_0 \cdot R_s. \tag{33}$$

As its name suggests, the geometric factor again becomes independent of the losses in the cavity, since $Q_0$ is inversely proportional to the surface resistance $R_s$. With the orders of magnitude of $Q_0$ and $R_s$ given above, $G$ is typically in the range of 100 Ω.

## 3.9 Equivalent circuit

The operating mode of a cavity can be conveniently represented and calculated using an equivalent circuit, in which the accelerating voltage and the beam current appear directly. Figure 11 shows this equivalent circuit. The power generator on the left is included as an ideal current source shunted by its inner resistance, the cavity itself is represented by an *RLC* resonance circuit, and the beam can be included as an ideal current source again. The inner resistance of the source is marked as $R/\beta$, with the coupling factor $\beta$ introduced above, and now this name should become clear: if $\beta = 1$, the inner resistance of the source is equal to the shunt impedance $R$, i.e., the power generator is matched to the cavity. For $\beta > 1$ the generator is said to be 'over-coupled', for $\beta < 1$ it is 'under-coupled'.

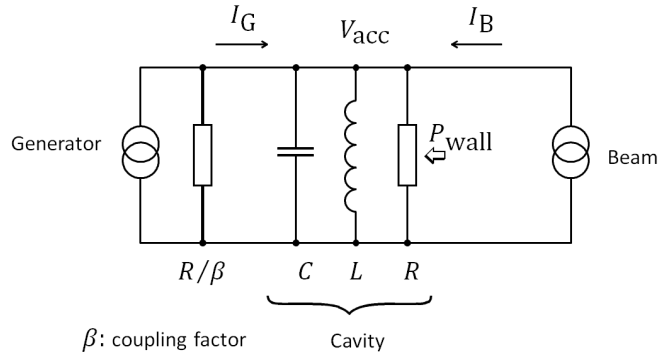

**Fig. 11:** Equivalent circuit of a resonant mode of a cavity

The equivalent circuit is governed by the equation

$$Y(\omega) \cdot V_{\text{acc}} = \left(\frac{1}{R} + j\omega C + \frac{1}{j\omega L}\right) \cdot V_{\text{acc}} = \frac{1}{R}\left(1 + j\, Q_0 \left(\frac{\omega}{\omega_0} - \frac{\omega_0}{\omega}\right)\right) \cdot V_{\text{acc}} = I_{\text{G}} + I_{\text{B}}, \quad (34)$$

where $R$ is the shunt impedance. When the right-hand side of Eq. (34) is set to zero (i.e., the cavity is disconnected from the generator and sees no beam), the remaining homogeneous equation can be understood as a non-standard eigenvalue problem in one dimension. A non-trivial solution exists for $\omega$ equal to the complex eigenfrequency (eigenvalue):

$$\omega_{\text{r}} = \omega_0 \sqrt{1 - \frac{1}{4Q_0^2}} + j\frac{\omega_0}{2Q_0}. \quad (35)$$

This eigensolution describes a damped oscillation with fields decaying as $\exp\left(-\frac{\omega_0 t}{2Q}\right)$, hence the stored energy decays at twice this rate. This eigensolution can be equated to the solution described in Eq. (20), except that here we have used the knowledge of the complex eigenfrequency, with the imaginary part describing the losses.

The circuit elements can be related to the characteristic quantities describing the cavity mode to get:

$$\begin{aligned} L &= \frac{R}{Q_0 \omega_0}, \quad C = \frac{Q_0}{R\omega_0}, \\ \omega_0 &= \frac{1}{\sqrt{LC}}, \quad \frac{R}{Q} = \sqrt{\frac{L}{C}}, \quad Q_0 = \frac{R}{\sqrt{L/Q}}. \end{aligned} \quad (36)$$

### 3.9.1 Resonant behaviour

When driving the cavity from the generator with a current at $\omega$, we can take advantage of the resonance behaviour of $Z(\omega) = 1/Y(\omega)$ (see Fig. 12) to create a large voltage $V_{\text{acc}}$ with a relatively small generator current/power.

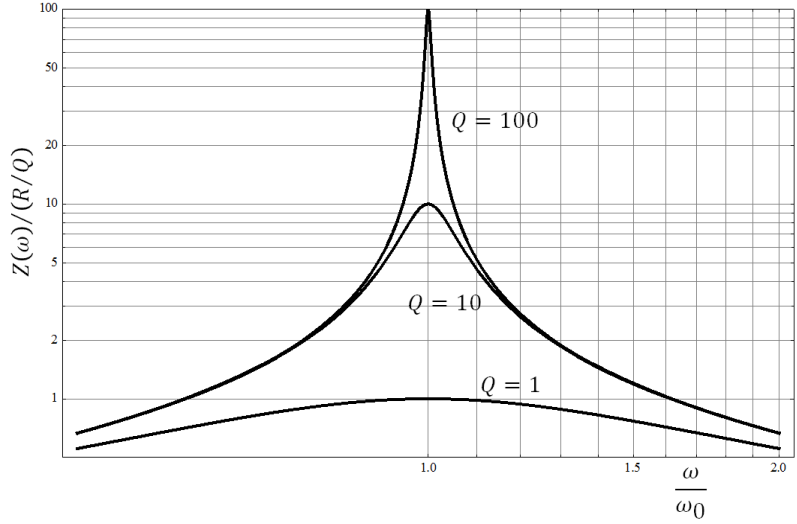

**Fig. 12:** Resonant behaviour of the cavity impedance for different values of $Q$

A note is appropriate here on the different definitions of the impedances currently in practical use and in the literature. We are using here the 'circuit' definition of the shunt impedance. We could have used the 'linac' definition, which is a factor of 2 higher, by replacing all occurrences of $R$ in the preceding equations with $R/2$. It should be clear however that this is just a definition. We try to be consistent within this text. Usually the 'circuit' definition is often used for synchrotrons, whereas 'linac ohms' are used in linacs (as the name suggests).

### 3.9.2 *Maximizing the shunt impedance*

In order to achieve a maximum accelerating voltage with a minimum power loss, it follows from Eq. (32) that the quantity $R$ should be maximized. One way to increase $R$ is to maximize $Q_0$, as, e.g., with superconducting cavities. The other way is to increase $R/Q$, i.e., to optimize the geometry in order to minimize the stored energy for a given accelerating voltage. A good example for the latter optimization was in fact already given in Fig. 9. The shape follows from an optimization of the shunt impedance; the remarkable features are (a) the 'nose cone', i.e., the nose-shaped profile which shortens the gap, concentrating the electric field near the beam passage but which keeps the transit time factor large, and (b) the round, almost spherical, external shape in the region where the magnetic field energy is stored, which allows the surface current (that leads to losses) to take the shortest possible path over the largest possible area.

In Fig. 9 we compared the obtained parameters for $R$ and $Q_0$ to an 'optimized pillbox' case. The $R/Q$ of a pillbox cavity is given in Eq. (26); its $Q_0$ can be calculated analytically to give

$$Q_{0,\text{pillbox}} = \frac{\sqrt{2\,a\,\eta\,\sigma\chi_{01}}}{2\left(1+\frac{a}{\ell}\right)}, \tag{37}$$

where $\sigma$ is the wall conductivity, which was set to $\sigma = 5.8 \cdot 10^7$ S/m for the numbers given in Fig. 9.

### 3.10 Beam loading

Beam loading denotes the loading of the cavity, and consequently of the generator, by the beam. The equivalent circuit in Fig. 11 already included the beam current. This allows us to estimate which fraction of the generator power will actually be absorbed by the beam. This power is given by $-\frac{1}{2}\Re\{V_{\text{acc}}I_B^*\}$ (we use the convention to count the power as positive when delivered *from* the beam).

When accelerating a beam, this power is negative and should be large for high efficiency. In other words, beam loading is desired, since this power is the power delivered to the beam.

On the other hand, beam loading also has its adverse effects: when optimizing for a high $R$, as in Section 3.9.2, a relatively small beam current can induce a relatively high gap voltage; in the equivalent circuit description, it does not make a difference whether the cavity is excited with a current $I_G$ from the generator or with a current $I_B$ from the beam (except that the generator output impedance, $R/\beta$, which appears in parallel to the shunt impedance, results in $R/(1+\beta)$ seen from the beam). In other words: a good accelerator is at the same time a good decelerator—the beam–field interaction goes both ways. The difference is that the cavity is tuned to the single frequency $\omega_0$ to optimize excitation at $\omega = \omega_0$ from the generator, whereas from the beam side we are worried about a wider frequency range. Another difference of course is that a high peak of $|Z(\omega)|$ seen from the generator is desired, whereas it is undesired as seen from the beam!

### 3.10.1 Modal loss factor

It is instructive to investigate the beam loading in the time domain: the passage of a point charge $q$ at time 0 through the initially empty (single-mode) cavity is equivalent to deposition of this charge on the equivalent capacitor $C = Q/(R\omega_0)$. The induced voltage is thus $q/C$, and the energy lost by the particle is

$$\Delta W = \frac{q^2}{2C} = \frac{\omega_0}{2}\frac{R}{Q}q^2 \equiv k_{\text{loss}}q^2. \tag{38}$$

This equation defines the (modal) loss factor $k_{\text{loss}}$.

After the passage of the charge, the voltage in the cavity will 'ring' as

$$V(t) = \begin{cases} 0 & \text{if } t < 0 \\ 2k_{\text{loss}}q\, e^{-\frac{\omega_0}{2Q_L}t} \cos\left(\omega_0\sqrt{1-\frac{1}{4Q_L^2}}\,t\right) & \text{if } t > 0 \end{cases} \tag{39}$$

according to Eq. (36), but with the loaded $Q$, $Q_L = Q/(1+\beta)$. This ringing is illustrated in Fig. 13.

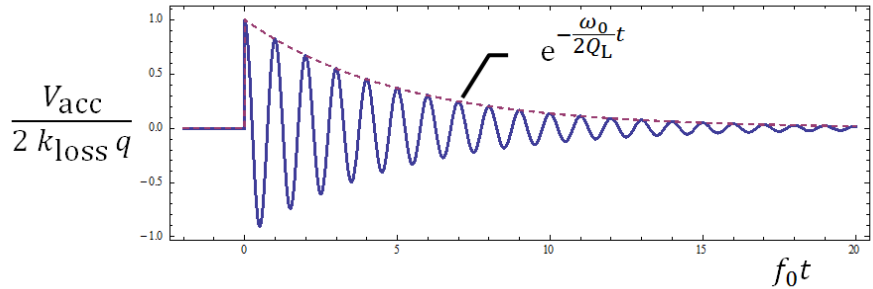

**Fig. 13:** Single-mode excitation by the passage of a point charge with assumed $Q_L = 15$

From the frequency domain calculation of a single mode, the modal loss factor is readily calculated using Eqs. (25) and (38) and becomes

$$k_{\text{loss}} = \frac{|V_{\text{acc}}|^2}{4W} = \frac{\omega_0}{2}\frac{R}{Q}. \tag{40}$$

It may appear surprising that the loss factor contains $R/Q$, but not the shunt impedance. In other words: the energy lost by a single charge in the cavity does not depend on $Q$ but only on the cavity geometry. This can be explained as follows: the loss of energy of a passing point charge happens fast on the time scale of the ringing—when the $Q$ of the cavity starts to play a role, the charge will be long gone. This is of course entirely different if the next particle follows one RF period behind the first, since it will not find an empty cavity.

### 3.10.2 Reactive and transient beam loading

In a synchrotron that is filled by batches (like the CERN SPS), the ring is often only partially filled, making the beam current discontinuous, leading to transients of the beam current $I_B$ (see Fig. 11). When the synchrotron's magnetic cycle has a flat top, i.e., when the beam is not accelerated or decelerated, longitudinal beam dynamics will ensure that in the presence of an accelerating voltage the bunches will form at its zero crossings, so, in the equivalent circuit of Fig. 11, $I_B$ will be 90° out of phase w.r.t $V_{acc}$, i.e., the beam loading will be purely reactive. Without going into this problem too deeply, let us consider transient reactive beam loading in the following. We assume we are above transition, where particles with excess momentum arrive late at the gap. These particles need to be decelerated, so the stable phase is the zero crossing from positive to negative gap voltage, i.e., the beam current is lagging the accelerating voltage by 90°. Seen from the amplifier, this is equivalent to an inductive susceptance. In principle, this could be entirely compensated by a parallel capacitance, in other words by detuning $\omega_0$ slightly below the operating frequency $\omega$. The amount of this detuning by the beam current is given by

$$-\Delta \equiv \frac{\omega_0 - \omega}{\omega_0} = -\frac{1}{2}\left(\frac{R}{Q}\right)\frac{|I_B|}{V_{acc}}. \tag{41}$$

Smart methods to address the problem of transient beam loading have been developed, including the correct choice of tuning $\omega_0$ and the invention of feedback loops that can rapidly follow the beam transients, but here it may suffice to state that relatively low values of $R/Q$ will help alleviate the problem.

Below transition the sign is different, and the beam will appear as a capacitive rather than an inductive susceptance, but in both cases the beam loading is reactive.

### 3.10.3 Wake fields

It should be noted in passing that the field left behind in a cavity by a passing charge will in general have an effect on trailing charges. If we assume the existence of a second charge following at a distance behind the leading charge, the integrated force on this second charge is referred to as the 'wake potential', so the function plotted in Fig. 13 is also known as the wake function (for a single-mode cavity). In the more general case, a number of modes will be excited and the wake function will be the superposition of the wake functions for the individual modes.

### 3.10.4 RF to beam efficiency

High RF to beam efficiency necessarily means high fundamental beam loading. In steady state, this efficiency is simply

$$\eta = \frac{P_{beam}}{P + P_{beam}}, \tag{42}$$

where $P_{beam} = -\frac{1}{2}\Re\{V_{acc} I_B^*\}$ is the power delivered to the beam, and $P = \frac{|V_{acc}|^2}{2R}$ is the power lost in the cavity wall according to Eq. (32). Assuming for simplicity that the beam current and accelerating voltage are in exact anti-phase (acceleration on the crest of the RF), this can be simplified:

$$\eta = \frac{1}{1 + \frac{V_{acc}}{R|I_B|}}. \tag{43}$$

With a given beam current, minimizing $V_{\text{acc}}/R$ will maximize the efficiency. Obviously, if a large accelerating voltage is desired, the only method to increase efficiency is to increase $R$ further, which, since there is a physical limit for $R/Q$, is possible only by a further increase of $Q$ (e.g., using superconducting cavities) or when using multi-gap accelerating structures.

## 3.11 Higher order modes

It is convenient to describe the behaviour of the cavity near its resonance as a lumped element circuit. The boundary conditions of the real cavity allow however for a multitude of eigensolutions, and every single one of them can be represented by an equivalent circuit like the one in Fig. 11, except that they are generally not driven (or controlled) by a generator, only by the beam. Especially for high intensity beams, these resonances can have a detrimental effect on beam stability, in particular if they happen to have a high $(R/Q)Q_{\text{L}}$. We illustrate HOMs in the following by the example of the CERN 80 MHz cavity, installed in the PS in 1997 and used to prepare the LHC bunches. Figure 14 shows this cavity in a sketch and in a photo.

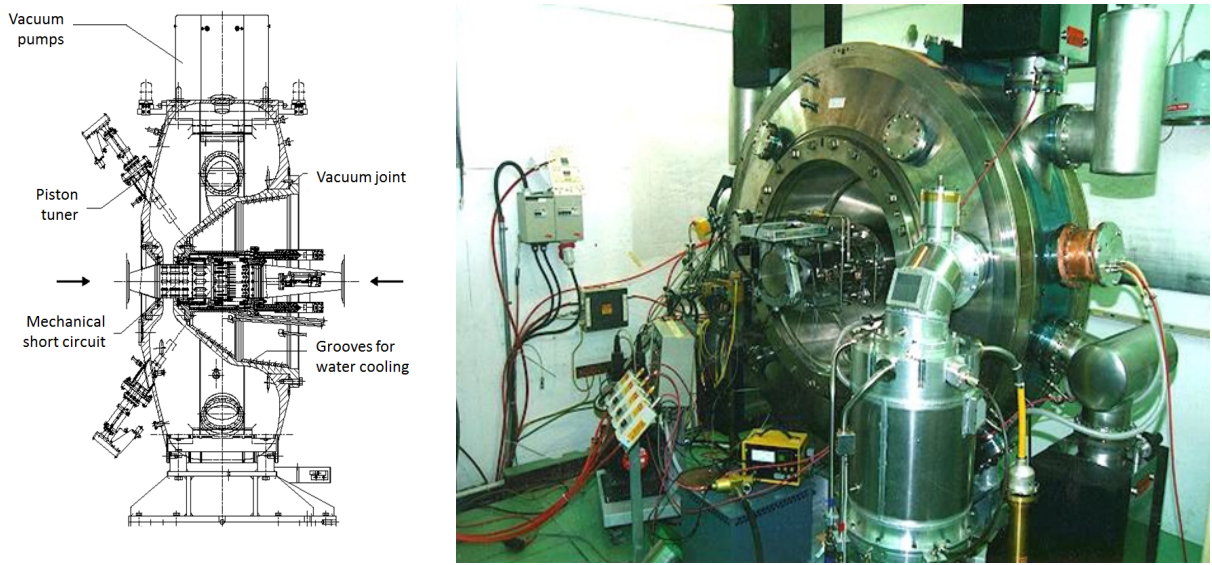

**Fig. 14:** The CERN/PS 80 MHz cavity used to illustrate HOM damping

The analysis of the HOM content of this cavity results in a large number of modes, the first 18 of which are depicted in Fig. 15, showing in colour code the intensity of the electric field, $|\vec{E}|$. In addition to their individual eigenfrequencies, these modes are of course characterized by their individual field distribution and consequently $R/Q$; some of them turn out to be more dangerous to the beam than others. It remains an important task to reduce their effect on the beam—after the field analysis this was done in the case of this cavity by identifying suitable locations on the outer cavity wall where feed-throughs could be placed. Antennas (or couplers) were connected via those feed-throughs' external damping resistors to damp the HOMs to a sufficient level. Often filters are inserted between the cavity and the HOM damper in order to prevent unwanted damping of the fundamental mode.

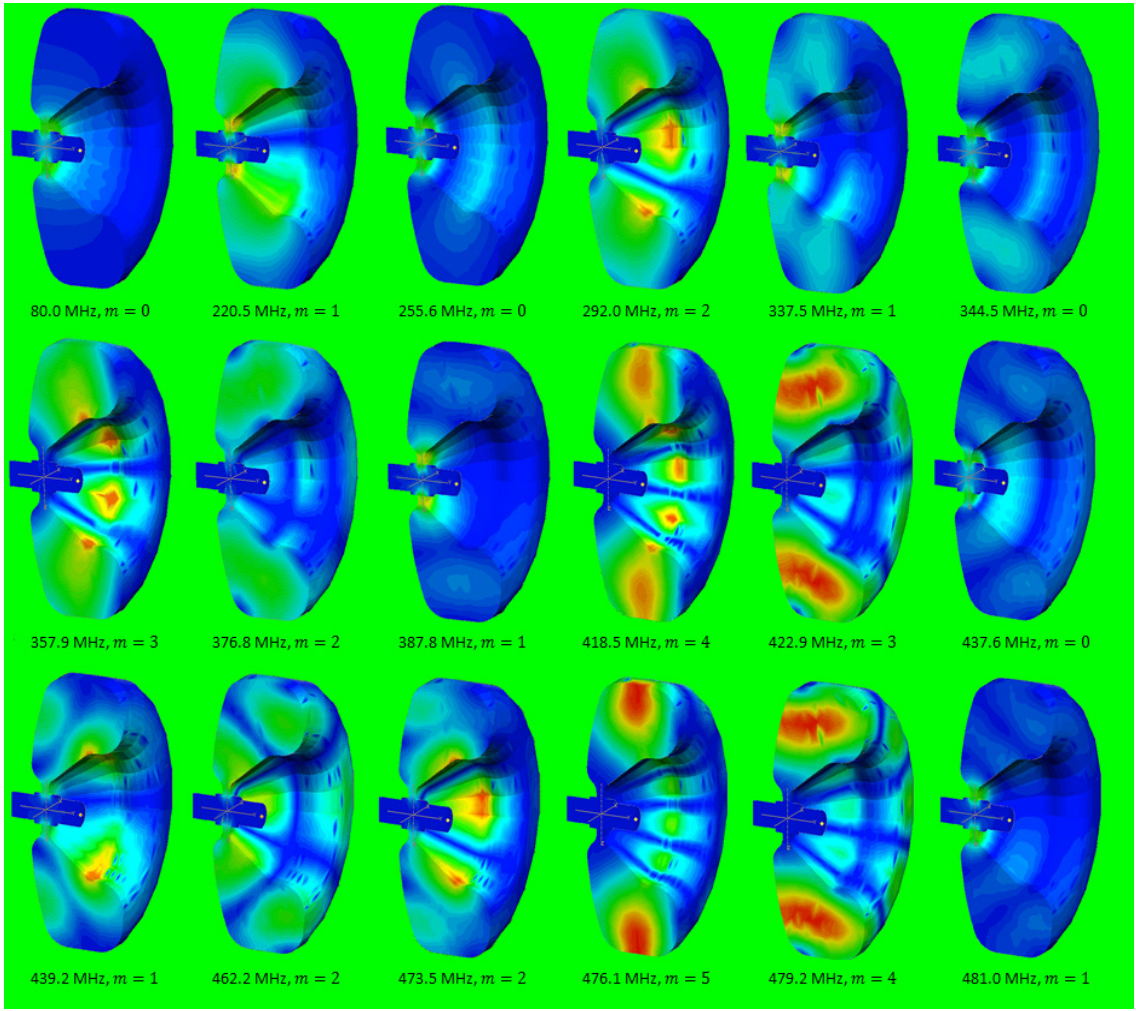

**Fig. 15:** The first 18 modes in the 80 MHz cavity shown in Fig. 14

The coupling to the external HOM dampers can be understood by generalizing the equivalent circuit shown in Fig. 11 by adding an ideal transformer—the resulting circuit is sketched in Fig. 16.

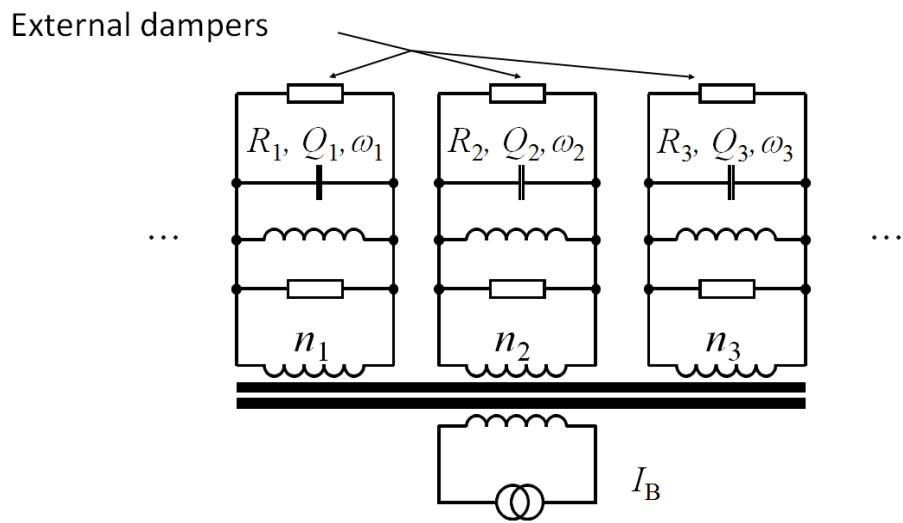

**Fig. 16:** The extension of the equivalent circuit of Fig. 11 to include HOMs

The transformer describes the HOM couplers or antennas, the different step-up ratios account for the different field distributions, and the resistors damp the modes by reducing their $Q_L$.

## 4 Multi-gap cavities

### 4.1 How many gaps?

As we have seen, the shunt impedance of a normal-conducting single-gap cavity cannot be significantly increased—the simple pillbox provides a good reference, and careful optimization could give slightly larger values. Superconducting RF of course allows a way out; another method is the use of multiple gaps, as will be discussed here.

To introduce multi-gap cavities let us see what happens if we just increase the number of gaps, keeping the total power constant: consider $n$ single-gap cavities with a shunt impedance $R$, as sketched in Fig. 17. The available power $P$ is split into equal parts and evenly distributed to the $n$ cavities. According to Eq. (32), each cavity will produce an accelerating voltage $\sqrt{2R(P/n)}$, so, with the correct phasing of the RF, the total voltage will be just the sum, $|V_{\text{acc}}| = \sqrt{2(nR)P}$. If we now consider the assembly consisting of the $n$ cavities and the power splitter as a single cavity with $n$ gaps, we notice that this new cavity has an effective shunt impedance $nR$; this is a significant increase. Consequently, by simply multiplying the number of gaps we can make much more efficient use of the available RF power to generate very large accelerating voltages.

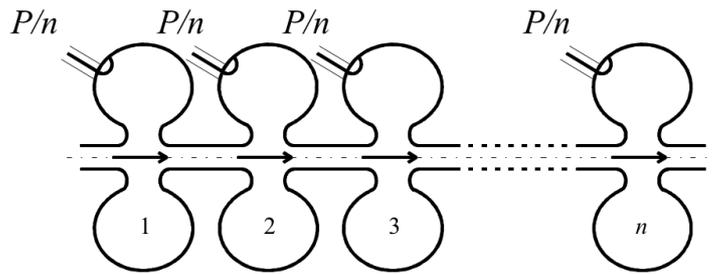

**Fig. 17:** Distributing a given power $P$ to $n$ cavities

Instead of using $n$ individual power couplers and a large power splitter, much more elegant ways of distributing the available RF power to many gaps have been invented—one is to combine the individual gaps in one vacuum vessel, and we can in fact use this vacuum vessel itself as a distributed power splitter, which leads to standing-wave or travelling-wave cavities.

### 4.2 Travelling-wave cavities

We introduced travelling waves when we derived waveguide modes from homogeneous plane waves. But we have also seen in Section 3.2 that the travelling wave in a hollow vacuum waveguide always has an axial phase velocity $v_{\varphi,z} > c$. A homogeneous plane wave, on the other hand, which has $v_{\varphi,z} = c$ in the direction of its propagation, will have no electric-field component in this direction. The purpose of travelling-wave structures thus can be understood as slowing down the phase velocity of the axial wave such that it becomes equal to the particle speed $\beta c$. Travelling-wave structures are often designed for $\beta = 1$, i.e., for electron linacs or more generally for particles with large enough $\gamma$. In this case, accelerating structures are designed as periodic structures with cell length $d$ and a phase advance of $\varphi = 2\pi d/\lambda$ per cell; this allows for continued acceleration over large distances.

In a travelling-wave structure, the RF power is fed via a power coupler into one end of the cavity (see Fig. 18); it flows (travels) through the cavity, typically (but not necessarily) in the same direction as the beam, creating an accelerating voltage at every gap. An output coupler at the far end of the cavity is connected to a matched power load. If no beam is present, the input power reduced by the cavity losses goes to the power load, where it is dissipated. In the presence of a large beam current, however, a large fraction of the forward travelling power can be transferred to the beam, such that a much smaller power comes through the output coupler.

Methods to analyse periodic structures have been developed for the analysis of periodic crystal lattices, looking at the wave function of electrons in a semiconductor for example. The result of such an analysis is the *Brillouin* diagram (Fig. 19). Only the first *Brillouin zone* $[0, \pi]$ for the phase advance $\varphi$ per cell is shown—the diagram should be considered mirrored at $\varphi = 0$ and periodically repeated with period $2\pi$. The solid line indicates a 'mode' of the periodic structure. It can be compared to the waveguide dispersion shown in Fig. 3, but with swapped abscissa and ordinate. As can be seen in Fig. 19, there is a lowest cutoff frequency, as in a homogeneous waveguide, below which no wave propagation is possible. At this cutoff, the dispersion curve starts with zero slope (zero group velocity) and an initial curvature very similar to that of the homogeneous waveguide. With increasing frequency, however, the dispersion curve bends down to reach another cutoff frequency when the phase advance of $\pi$ per cell is reached. A characteristic stop-band follows—a frequency range in which, as for below cutoff, no propagation is possible. In the pass-band, the curvature of the dispersion curve reaches a point when the phase velocity of the forward wave is equal to the speed of light, indicated in Fig. 19 by a dashed line. This intersection is the design point; it has a specified phase advance per cell (in this case $2\pi/3$) for the operating frequency. At this point a particle with the speed of light is synchronous with the mode of the periodic structure and thus sees the correct RF phase in every cell—one can imagine the particle 'surfing' on this forward-travelling wave.

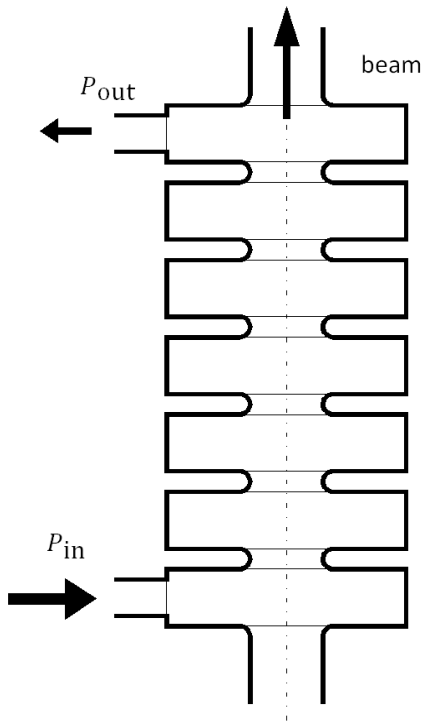 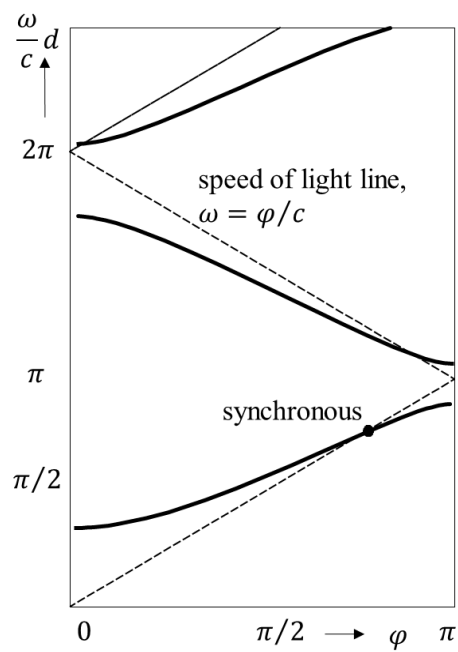

**Fig. 18:** Sketch of a travelling wave structure    **Fig. 19:** A Brillouin diagram for a periodic structure

Figure 20 shows a cut-away view of simple multi-cell travelling-wave cavities made from brazed copper discs.

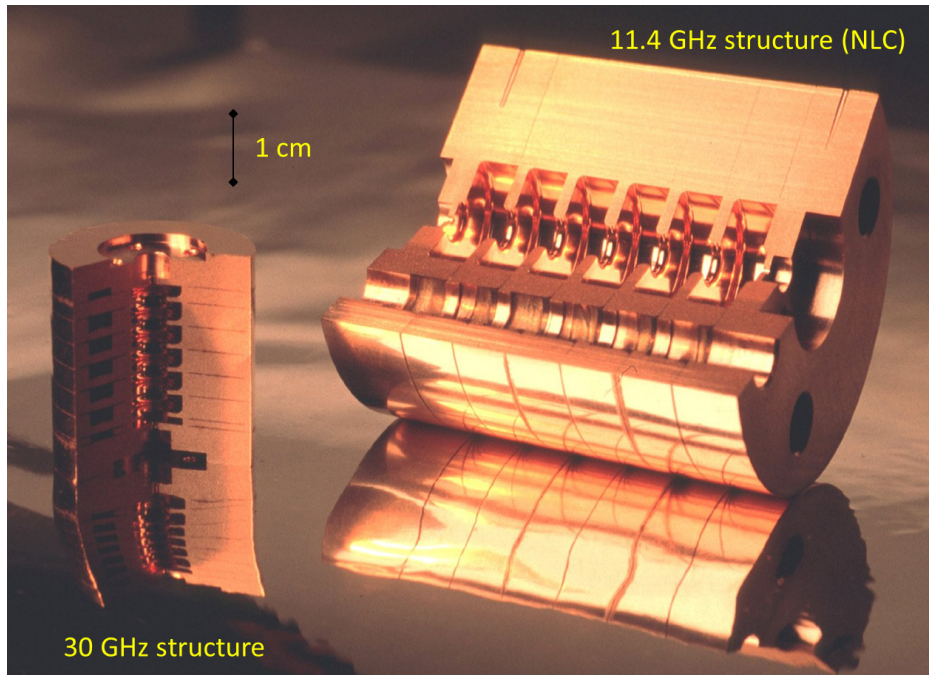

**Fig. 20:** Cut-away view of travelling-wave accelerating structures, fabricated from brazed Cu discs

### 4.3 Standing-wave cavities

When the matched load at the output port (see Fig. 18) of a travelling-wave structure is replaced by a short circuit, a standing wave will build up. Since the backward-travelling wave does not intersect with the speed of light line, this reflected wave will have no net contribution to the acceleration; this standing wave would work, but it would not be a good design. This is different for the phase advances of $0$ ($= 2\pi$) and $\pi$ per cell, where forward and backward waves can be considered 'degenerate' to form a standing wave, which makes these points well suited for a cavity. They feature zero slope of the dispersion curve, which means that at these points no power is travelling in the structure, the group velocity is zero, and the energy is stored. As an example, Fig. 21 shows the electric-field lines of a $\pi$-mode, which should be designed to have a cell length of $\beta\lambda/2$ for a particle speed $\beta c$.

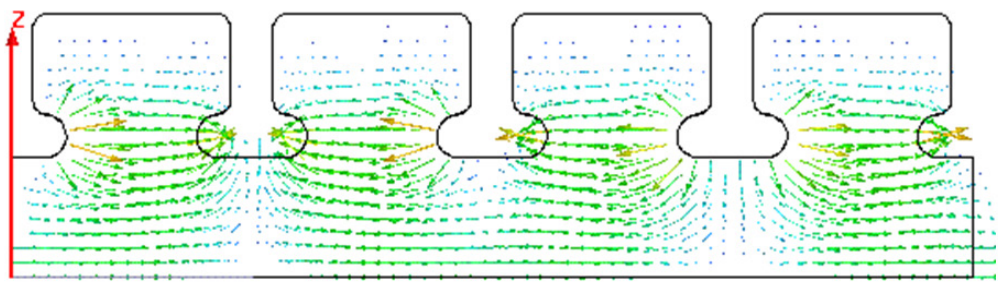

**Fig. 21:** Electric field lines of the $\pi$-mode of a periodic structure. This standing-wave cavity has a cell length of $\beta\lambda/2 = \pi\beta c/\omega$ for a particle speed $\beta c$.

The example of the $\pi$-mode structure was chosen here since it is widely used; in particular, virtually all superconducting cavities are designed to operate in $\pi$-mode. As illustrated in Fig. 22, the cells of superconducting multi-cell cavities have a characteristic shape constructed from elliptic arcs. This feature results from optimization of the field distribution (minimizing the surface electric field), the suppression of the multipactor, and the possible reduction of HOMs (with a large beam pipe that allows the HOMs to propagate out). A practical implementation of a four-cell superconducting cavity is shown on the right.

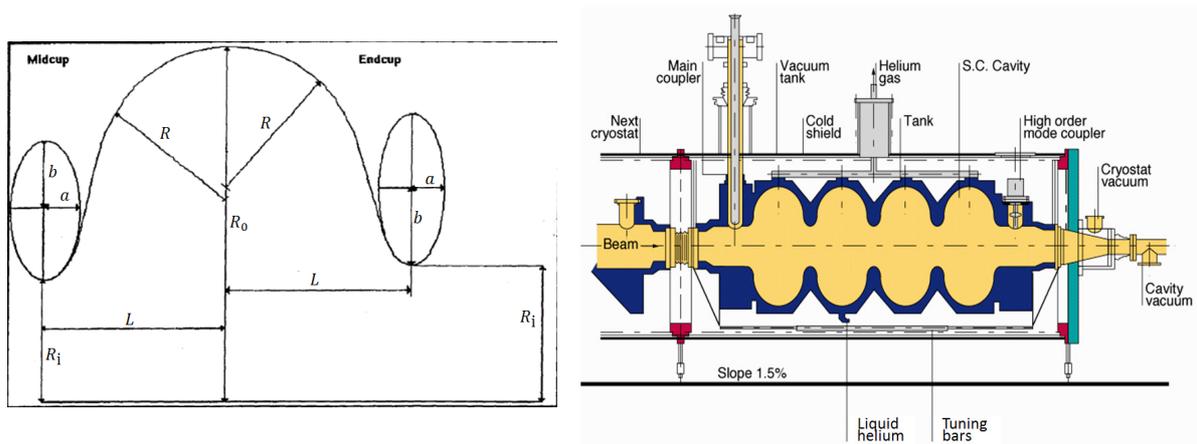

**Fig. 22:** Left: characteristic 'elliptical' cell shape used in superconducting cavities [5]; right: example of a four-cell superconducting cavity.

## References

[1] http://laacg.lanl.gov/laacg/services/download_sf.phtml

[2] http://www.ansys.com/Products/Simulation+Technology/Electronics/Signal+Integrity/ANSYS+HFSS

[3] https://www.cst.com/

[4] http://www.gdfidl.de/

[5] D. Proch, http://accelconf.web.cern.ch/accelconf/SRF93/papers/srf93g01.pdf